\documentclass[reprint,aps,pre,amssymb]{revtex4-1}

\usepackage[T1]{fontenc}
\usepackage{ucs}
\usepackage[utf8x]{inputenc}

\usepackage[reqno,intlimits]{amsmath}
\usepackage{amssymb}
\usepackage[final]{graphicx}
\usepackage{float}
\usepackage{mathrsfs}
\usepackage{dsfont}

\usepackage[usenames,dvipsnames]{color}

\newcommand{\pb}[1]{
  \parbox[0pt][#1][c]{0cm}{}
}

\newcommand{\trace}[1]{
  \mathrm{trace}\left\{ #1 \right\}
}

\renewcommand{\vec}[1]{
  \mathbf{#1}
}

\begin{document}
  \allowdisplaybreaks
  \DeclareGraphicsExtensions{.pdf}

  \title{Dynamical localization and eigenstate localization in trap models}
  \date{\today}
  \author{Franziska Flegel and Igor M. Sokolov}
  \affiliation{%
  Institut f\"{u}r Physik, Humboldt-Universit\"{a}t zu Berlin, D-12489 Berlin, Germany%
  }
  \begin{abstract}
  The one-dimensional random trap model with a power-law distribution of mean sojourn times exhibits a phenomenon of dynamical localization
  in the case where diffusion is anomalous: The probability to find two independent walkers at the same site, as given by the participation ratio,
  stays constant and high in a broad domain of intermediate times.
  This phenomenon is absent in dimensions two and higher. In finite lattices of all dimensions the participation ratio finally equilibrates to a
  different final value. We numerically investigate two-particle properties in a random trap model in one and in three dimensions,
  using a method based on spectral decomposition of the transition rate matrix. The method delivers a very effective computational scheme producing numerically exact results
  for the averages over thermal histories and initial conditions in a given landscape realization. Only a single averaging procedure over disorder realizations is necessary.
  The behavior of the 
participation ratio is compared to other measures of localization, as for example to the states' gyration radius, according to which the dynamically localized states are extended.
This means that although the particles are found at the same site with a high probability, the typical distance between them grows. Moreover the final equilibrium state is extended
both with respect to its gyration radius and to its Lyapunov exponent. In addition, we show that the phenomenon of dynamical localization is only marginally connected with the spectrum
of the transition rate matrix, and is dominated by the properties of its eigenfunctions which differ significantly in dimensions one and three.   
  \end{abstract}

  \maketitle
  
  \section{Introduction}
 
  The random trap model together with its close relative, the barrier model, can be applied to a variety of physical problems \cite{Haus1987} which are related to random walks in disordered media, like properties of a photocurrent in amorphous solids \cite{Scher1975}, 
 or have an equivalent mathematical representation, like the behavior of resistor and capacitance networks \cite{Bouchaud1990} or lattice vibrations of harmonic chains \cite{Alexander1981,Dyson1953}.
   It has also become prominent as a toy model for the phase space dynamics of spin glasses \cite{Bouchaud1992}.
  
  The random trap model on a regular lattice
  consists of a trapping landscape $\{ E_k \}$ of energy wells located on the lattice's sites $k$. The $E_k$ are iid random variables which in our case
  have the common density
  \begin{align}
     \rho_E (\epsilon) &= \frac{1}{kT_g} ~\exp\left(-\frac{\epsilon}{kT_g}\right),
  \end{align}
  with $T_g$ being the characteristic temperature. According to the Van't Hoff-Arrhenius law \cite{Haenggi1990}, at a temperature $T$ these trapping potentials correspond to
  \emph{mean waiting times} $\tau_k = \tau_0 \exp\left( \frac{E_k}{kT} \right)$, where we put $\tau_0 = 1$ in what follows.
  The density of the mean waiting times $\tau_k$ is
  \begin{align}
    \rho_\tau (t) &= \alpha t^{-1-\alpha}\qquad (t \geq 1),
    \label{equ:rhotau}
  \end{align}
  with $\alpha = T/T_g$.
  In what follows, we consider lattices in dimensions $D = 1$ and $D = 3$.
  In 1D the particle's dynamics on the lattice is described by a continuous-time Markov chain with transition rates
  \begin{align}
  w_{k\rightarrow l} &= \begin{cases}
                          -1/\tau_k, &\text{if~} k = l,\\
                          1/(2D\tau_k), &\text{if~} k,l \text{~neighbors},\\
                          0, &\text{else},
                        \end{cases}
                        \label{equ:GeneralRates}
  \end{align}
  with $2D$ being the number of neighbors of site $k$ and $\tau_k$ denoting the mean waiting time associated with the site. If the first moment of the distribution, Eq.(\ref{equ:rhotau}), diverges, which is the case for $0 < \alpha < 1$,
  the dynamics of the system starts to show highly nontrivial behavior (as exemplified by anomalous diffusion in such models).  
  Of course, one could also define a mixture of the trap and barrier model, or generalize the model as in \cite{BenArous2005}.
  
  The model above describes the behavior of random walkers who, while exploring the lattice,
  get trapped in the potential wells. Once trapped at site $k$, they have to wait for a time $\tau$ (actual waiting time) with pdf
  \begin{align}
   \psi_k (\tau) &= \frac{1}{\tau_k} \exp\left(-\frac{\tau}{\tau_k}\right),
   \label{equ:exponentialpdf}
  \end{align}
  containing the mean waiting time $\tau_k$ at a site $k$ as parameter.
  Note that these local means $\tau_k$ are distributed according to Eq. (\ref{equ:rhotau}) and the exponential waiting time density of Eq. (\ref{equ:exponentialpdf})
  is already conditioned on the fact that a random walker resides on a site with mean waiting time $\tau_k$.
  After the random walkers have waited long enough, they jump with equal probability to any of the recent site's neighbors.
  Thus, the spatial properties of each single random walker's trajectory are the same as in a simple random walk on the lattice. What is interesting, is the temporal evolution.
  The single-particle (one-time) quantities such as the mean squared displacement are well-understood, and can in high dimensions be described by the continuous-time
  random walk scheme; the behavior in 1D, which shares many properties with the CTRW behavior, is also well-investigated \cite{Bouchaud1990}. 
  The properties of the motion of two or more random walkers exploring the same energy landscape are much less understood.
  
\begin{figure}
   \includegraphics[width=\linewidth]{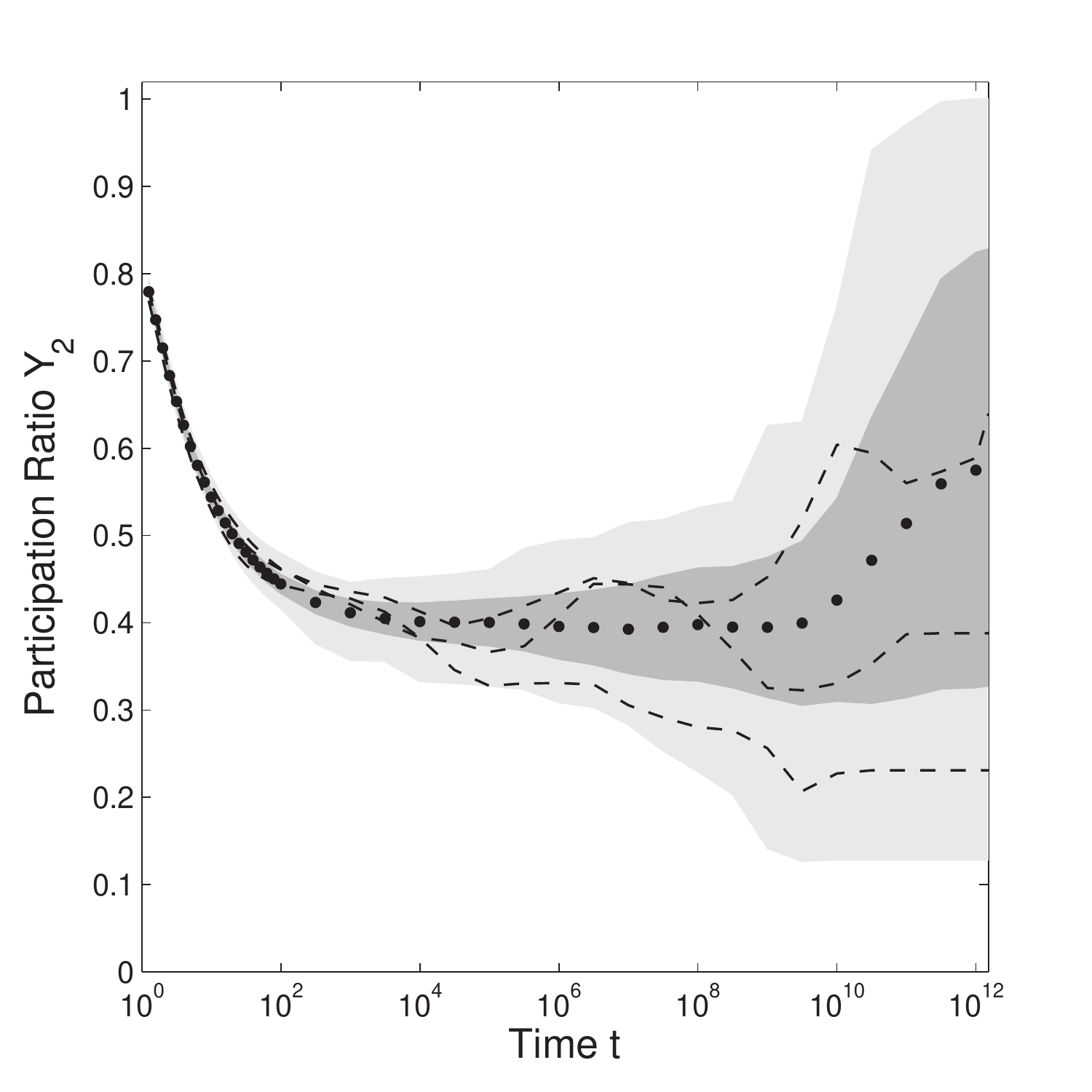}
   \caption{Temporal evolution of sample $\langle Y_2\rangle_{p_0}$ (dashed lines) and $\langle Y_2\rangle_{p_0, \tau}$ (dots) for $\alpha = 0.37$,
	      $N = 11^3$ and the disorder average over 100 realizations. The light gray area gives the whole range of $\langle Y_2\rangle_{p_0}$ of all 100 realizations and
	      the dark gray area depicts $\langle Y_2\rangle_{p_0,\tau} \pm \sigma$ with $\sigma$ being the standard deviation.}
   \label{fig:Y2p0av}
\end{figure}

  In what follows we concentrate on some of these multiparticle quantities, from which the (inverse) participation ratio $Y_2(t)$, which represents the probability
  that two random walkers who started at $t_0 = 0$ at the same site $k_0$ meet again at time $t$, was considered the most interesting.
  If the lattice is finite, the disorder-averaged long-time limit $\lim_{t \rightarrow \infty} \langle Y_2 (t)\rangle_\tau$
  converges to an equilibrium value $\langle Y_2 \rangle_\tau^{(eq)} = 1-\alpha$ \cite{Bertin2002,Derrida1997}, independent of the lattice's dimension $D$.
  The question about the non-equilibrium behavior of $\langle Y_2 \rangle_\tau$, however, is much more involved. It seems especially fascinating that in dimension one
  even for infinitely large lattices
  there exist a finite disorder-averaged long-time limit $\langle Y_2 \rangle_\tau^{(dyn)} >0$ with $\langle Y_2 \rangle_\tau^{(dyn)} \neq \langle Y_2 \rangle_\tau^{(eq)}$,
  the phenomenon called \emph{dynamical localization}.
  This manifests itself also in finite but large  one-dimensional lattices as a pronounced plateau
  in the temporal evolution of $\langle Y_2\rangle_\tau$ (see Fig. \ref{fig:Y2p0av}).
  The existence of this regime has been mathematically proven by Fontes \emph{et al.} \cite{Fontes2002} and
  was further investigated by Bertin and Bouchaud \cite{Bertin2002}. For lattices with dimension $D \geq 3$, Fontes \emph{et al.} \cite{Fontes1999} 
  proved that the dynamical localization is absent; it is conjectured that it is also absent in $D=2$. 
The phenomenon of dynamical localization is thus a property of one-dimensional lattices, i.e.:
\begin{align}
  \lim_{t\rightarrow \infty} \left( \lim_{N \rightarrow \infty} \langle Y_2\rangle_{p_0, \tau} \right) &= \begin{cases}
                                                                                                            0, &\text{if~} D \geq 2,\\
                                                                                                           \langle Y_2\rangle^{(dyn)}_\tau>0, &\text{if~} D = 1.
                                                                                                           \end{cases}
 \end{align}      
The order of the limits is crucial since otherwise the expression converges to the equilibrium value $1-\alpha$.

  The goal of this paper is to study the dynamical localization in somewhat more detail, and to connect it with the properties of the spectrum and of the eigenfunctions
  of the Laplacians defined by Eq. (\ref{equ:GeneralRates}) in the one- and three-dimensional lattices with periodic boundary conditions.
  This is done by extensive numerical simulations relying on the
  spectral representation of the dynamics. The method allows for a numerically exact calculation of the system's properties averaged over the initial 
  position of the walkers and their thermal histories (trajectories) for a given realization of traps, and for considering longer observation times than
  the direct Monte-Carlo simulations. Apart from investigating the behavior of $Y_2$, we consider some other measures of localization, as well as the properties of the
  eigenfunctions of the corresponding Laplacians, trying to hint onto the main properties of the system responsible for dynamical localization.
  
\section{The Laplacian matrix}

Let us consider a 1D ring of $N$ sites and a realization $\{ \tau_k \}$ of waiting times according to the prescribed waiting time distribution $\rho_\tau$ of Eq. (\ref{equ:rhotau}).
Let $\vec{p}(t) \in [0,1]^N$ be the vector whose components $\pi_k$ describe the probability to find the random walker at site $k$ at time $t$.
Then the participation ratio $Y_2$ is simply $\vec{p}^2 = \vec{p}^T \vec{p}$.

The temporal evolution of $\vec{p}$ is given by the matrix $W$ of the transition rates $w_{k\rightarrow l}$ which are
explicitly defined by Eq. (\ref{equ:GeneralRates}).
All eigenvalues of $W$ are nonpositive.
In what follows we will refer to the Laplacian matrix (Laplacian) $L = -W$, with $(L)_{lk}=-w_{k\rightarrow l}$ and nonnegative eigenvalues.

The master equation thus reads $\dot{\vec{p}} = -L\vec{p}$
and it leads to the temporal evolution of $\vec{p}$ in the form
\begin{align}
  \vec{p}(t) = \exp(-Lt) \vec{p}_0,
\end{align}
with $\vec{p}_0$ being the initial condition. In our case, the initial distribution $\vec{p}_0$ will always be completely localized, i.e. $\vec{p}_0^2 = 1$.

If we average over all these localized initial conditions, then $Y_2 = \vec{p}_0^T \exp(-L^T t) \exp(-L t) \vec{p}_0$ becomes the trace
\begin{align}
 \langle Y_2\rangle_{p_0} &= \frac{1}{N} ~\trace{e^{-L^T t} e^{-L t}}\label{equ:pTpSingularValue}\\
   &= \frac{1}{N} \sum_{j=1}^N \sigma_j(t),\nonumber    
\end{align}
with $\sigma_j(t)$ being the $j$-th eigenvalue of the time-dependent operator $\mathcal{O}(t) = e^{-L^T t} e^{-L t}$.
In what follows we will however use a more convenient representation and describe the process in the
basis of the eigenfunctions of the Laplacian $L$ where the spectral decomposition is time-independent.

If $\mathcal{L}$ is the Laplacian for a simple random walk on a ring of size $N$ (i.e. a Toeplitz matrix with $1$ on the main diagonal, minus one half on the two adjacent diagonals
and in the upper right and lower left corners, and all other elements equal to zero) and $S = \text{diag}\left(\frac{1}{\sqrt{\tau_1}}, \ldots, \frac{1}{\sqrt{\tau_N}}\right)$,
the Laplacian $L$ is given by
\begin{align}
  L = \mathcal{L}S^2.
\end{align}
The transpose of $L$ is
\begin{align}
  L^T &= S^2 \mathcal{L} = S^2 L S^{-2},
\end{align}
and the matrix
\begin{align}
  A &= SLS^{-1}
  \label{equ:A}
\end{align}
is symmetric and hence diagonalizable over $\mathbb{R}$. This diagonalizability is inherited to $L$ and $L^T$. 
We will denote the eigenvectors of $L$ by $\vec{X}_i$, of $A$ by $\vec{Q}_i$ and of $L^T$ by $\vec{Z}_i$.
Since $A$ is symmetric, we can choose the $\vec{Q}_i$ such that they form an orthonormal basis of $\mathbb{R}^N$ and define the eigenvectors of $L$ and $L^T$
such that for eigenvectors to the same eigenvalue $\lambda_i$ the following relation holds:
\begin{align}
  \vec{Z}_i &= S \vec{Q}_i = S^2 \vec{X}_i.
\end{align}
From this follows:
\begin{align}
  \delta_{ij} &= \vec{Q}_i^T \vec{Q}_j = (S \vec{X}_i)^T (S^{-1} \vec{Z}_j) = \vec{X}_i^T \vec{Z}_j,
  \label{equ:Orthogonality}
\end{align}
where we have used the fact that $S$ is diagonal. 
Furthermore, there is an equilibrium state of $L$, namely the eigenvector $\vec{X}_1 = (\tau_1, \tau_2, \ldots , \tau_N)^T/\sqrt{\sum_k\tau_k}$ to eigenvalue zero,
corresponding to the eigenvector $\vec{Z}_1 = (1, \ldots , 1)^T/\sqrt{\sum_k\tau_k}$ of $L^T$.
All other eigenvalues are strictly positive since $A = S^T\mathcal{L}S$ with $S$ being invertible and $\mathcal{L}$ being (symmetric) positive-semidefinite with
exactly one vanishing eigenvalue.
The eigenvalues can be ordered such that
\begin{align}
  0 = \lambda_1 < \lambda_2 \leq \lambda_3 \ldots.
\end{align}

\section{Participation ratio}
\label{sec:Y2Eig}

We want to express the temporal evolution of $Y_2 = \vec{p}^2$ in terms of eigenvectors and eigenvalues of
$L$ and $L^T$.
It is easily verified that
\begin{align}
  \vec{p} = \sum_i \left( \vec{p}_0^T \vec{Z}_i \right) \vec{X}_i e^{-\lambda_i t},
\end{align}
and thus
\begin{align}
   \vec{p}^T \vec{p}  &= \sum_{ij} \left( \vec{Z}_i^T \vec{p}_0 \vec{p}_0^T \vec{Z}_j \right) \left( \vec{X}_i^T \vec{X}_j\right) e^{-(\lambda_i + \lambda_j)t}.
   \label{equ:pTpInitial}
\end{align}
If we average over all initial conditions $\vec{p}_0$ fulfilling $\vec{p}_0^2 = 1$, then
\begin{align}
   \langle \vec{p}_0 \vec{p}_0^T\rangle_{p_0} &= \frac{\mathbb{I}_N}{N},
   \label{equ:p0average}
\end{align}
with $\mathbb{I}_N$ the $N \times N$ identity matrix. Note, that we have not yet applied any disorder average. Averaging Eq. (\ref{equ:pTpInitial}) over the initial conditions and 
applying Eq. (\ref{equ:p0average}) 
we obtain
\begin{align}
   \langle Y_2\rangle_{p_0} &= \frac{1}{N} \sum_{ij} G_{ij} e^{-(\lambda_i + \lambda_j)t}.
   \label{equ:pTpAveraged}
\end{align}
where the elements 
\begin{align}
  G_{ij} &= (\vec{Z}_i^T \vec{Z}_j)(\vec{X}_i^T \vec{X}_j)\label{equ:DefG}\pb{2em}\\
    &= (\vec{Q}_i^T S^2 \vec{Q}_j)(\vec{Q}_i^T S^{-2} \vec{Q}_j)\pb{2em}\nonumber,  
\end{align}
define a new matrix $G$ which is the element-wise product of the Gramian matrices corresponding
to the vector sets $\{ \vec{Z}_1, \vec{Z}_2, \ldots, \vec{Z}_N \}$ and $\{ \vec{X}_1, \vec{X}_2, \ldots, \vec{X}_N \}$.
Thus, the temporal development of $\langle Y_2\rangle_{p_0}$ depends on $G_{ij}$ in relation to the characteristic decay times $1/(\lambda_i + \lambda_j)$. 
As we proceed to show, these are the properties of $G$ which dominate the dynamical localization.

Taking the temporal limit of Eq. (\ref{equ:pTpAveraged}) in a finite system and averaging over disorder, we can also reproduce the finite-size equilibrium value $\langle Y_2 \rangle^{(eq)}_\tau$ 
as it was given in \cite{Bertin2002,Derrida1997}:
\begin{align}
  \left\langle \lim_{t\rightarrow \infty} \langle Y_2\rangle_{p_0} \right\rangle_\tau &= \frac{1}{N} \left\langle\left( \vec{Z}_1^T \vec{Z}_1 \right)
					  \left( \vec{X}_1^T \vec{X}_1\right)\right\rangle_\tau\nonumber\\
      &= \left\langle\frac{1}{\sum_k \tau_k} \left( \vec{X}_1^T \vec{X}_1\right)\right\rangle_\tau\nonumber\\
      &= \left\langle\frac{\sum_l \tau_l^2}{\left(\sum_k \tau_k\right)^2}\right\rangle_\tau\simeq 1-\alpha,
\end{align}
see \cite{Sokolov2010}.

Eq. (\ref{equ:pTpAveraged}) comes in very handy because it enables us to determine $\langle Y_2\rangle_{p_0}$ at any time $t$ without previously computing everything that happened before $t$.
Additionally, the temporal behavior of $\langle Y_2\rangle_{p_0}$
(which is still dependent on the disorder) can be simulated quite efficiently, i.e. without the statistical error attached to the average over initial
conditions and over single trajectories. Thus the matrix approach gives a useful numerical tool of investigating the problem of dynamical localization, the tool which will be 
continuously used in the present paper.
A nice consequence is that we are able to examine the temporal behavior of $\langle Y_2\rangle_{p_0}$ as averaged over initial conditions and trajectories
in a fixed disorder realization and compare it to the disorder-averaged behavior to evaluate the inter-sample differences, as is done in Fig. \ref{fig:Y2p0av}.

Fig. \ref{fig:Y2p0av} shows the behavior of $\langle Y_2\rangle_{p_0}$ in three different realizations of the random potential together with the result
$\langle Y_2 \rangle_{p_0, \tau}$ as following from averaging over 100 such realizations.
The pronounced plateau of $\langle Y_2 \rangle_{p_0, \tau}$ between $t \simeq 10^2$ and $10^9$ corresponds to the dynamical localization 
and up to now was only considered after applying the disorder average. We see that up to the crossover at $t \approx 10^8$ 
the $\langle Y_2 \rangle_{p_0}$ do not differ strongly from each other, so that the hints onto the dynamical localization are present in each single
realization of the random traps. Especially at the onset of the plateau at $t \approx 10^2$ all different $\langle Y_2 \rangle_{p_0}$
appear to follow the same pattern, independent on what equilibrium value of $\langle Y_2\rangle_\tau$ would be attained. Fig. \ref{fig:Y2p0av} shows that, even though the temporal 
evolution of the sample $\langle Y_2\rangle_{p_0}$ fluctuates around the disorder average $\langle Y_2\rangle_{p_0, \tau}$,
the effect of the dynamical localization is not an exclusive property of $\langle Y_2\rangle_{p_0, \tau}$ but does already appear
in the average over initial conditions and thermal histories.                                                                                                  

The analytical calculation of $Y_2$ 
in the non-equilibrium case pertinent to dynamical localization turned out to be quite difficult.
Bertin and Bouchaud \cite{Bertin2002} conjectured
\begin{align}
  \langle Y_2 \rangle_\tau^{(dyn)} &= \frac{2}{3}~\langle Y_2 \rangle_\tau^{(eq)},
  \label{equ:BBconjecture}
\end{align}
using a simple approximation
and presented simulations with infinite-time extrapolation that confirmed their conjecture. We find it worth mentioning
that according to their simulations \cite[Fig. 5]{Bertin2002} as well as in our attempts to reproduce their figure, the simulated data tend to be smaller than
Eq. (\ref{equ:BBconjecture}) predicts, whereas it is more likely that the fitting procedures overestimate the infinite-time value $\langle Y_2\rangle^{(dyn)}_{p_0, \tau}$.
Nevertheless, Eq. (\ref{equ:BBconjecture}) gives a simple and elegant approximation.

\section{Participation ratio and other localization measures}
\label{sec:dynloc}

\subsection{What does dynamical localization (not) mean?}
\label{subsec:locmeasures}

The localization of states was extensively studied for vibrational excitations and in quantum models, where different localization 
criteria were applied mostly to the eigenstates of the corresponding Hamiltonians; in a typical case all of them 
lead to the same kind of conclusions on whether the corresponding state is localized or not. 
Thus, the participation ratio quantifies to what extent the localized state is \textit{concentrated} on a small subset of all sites. 
The standard criterion for localization which is both theoretically transparent and easy for numerical implementation, especially in one dimension, is based on 
the evaluation of the Lyapunov exponent $\gamma$
which corresponds to the rate of exponential growth of the amplitude with the coordinate \cite{Ishii1973,Comtet2013,Thouless1972}, and determines the behavior of the 
state outside of its localization region. We will consider the behavior of the Lyapunov exponent later on in this paper.

One can also quantify the localization by the radius of gyration $R_G$ of the corresponding state,
which indicates the spatial \textit{spread} of the state, the size of the spatial domain which the particle or excitation is confined to.

\begin{figure}
  \includegraphics[width=\linewidth]{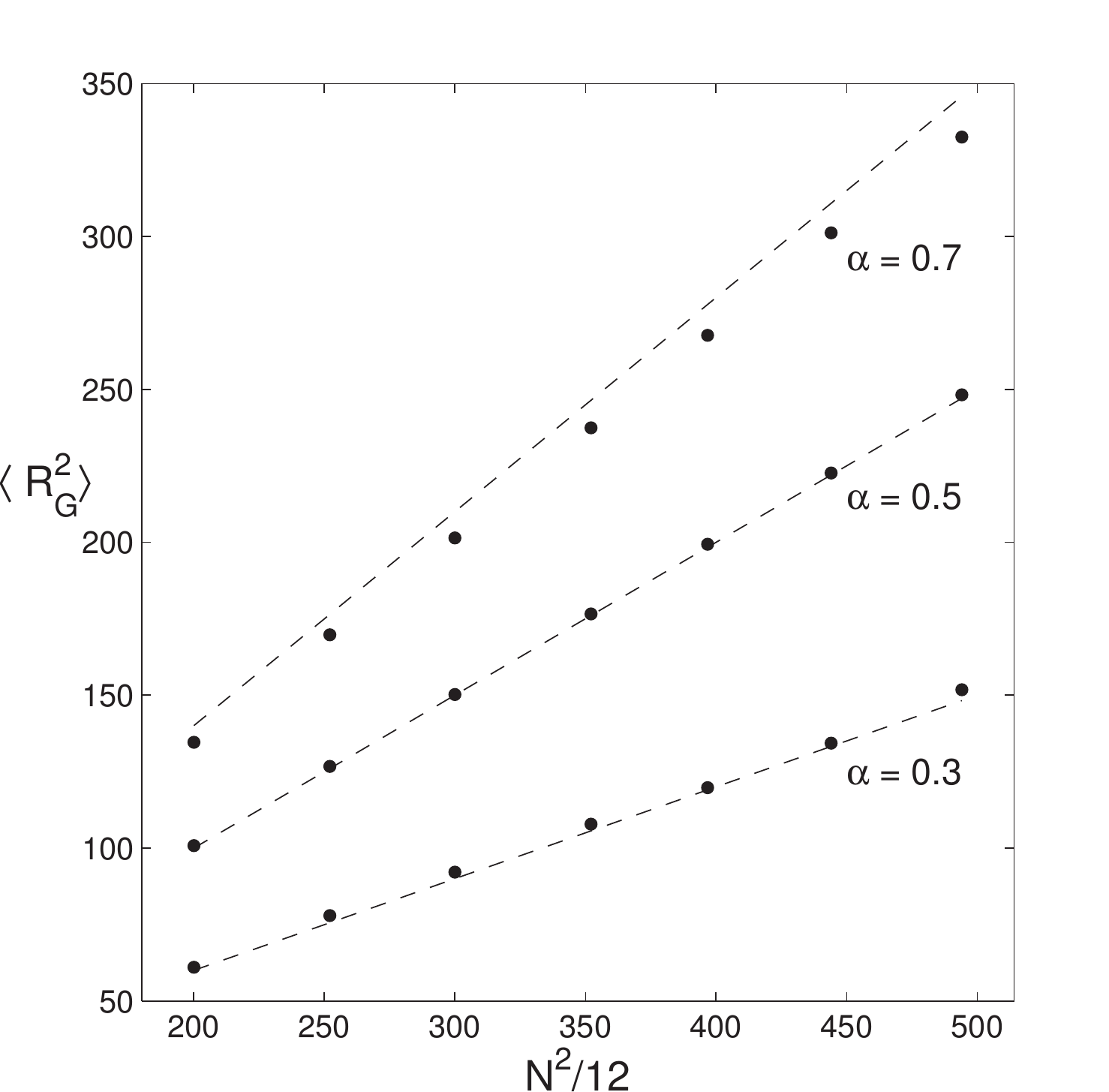}
  \caption{Equilibrium value $\langle R_G^2\rangle_\tau^{(eq)}$ of the radius of gyration in dependence of the system size. In agreement with Eq. (\ref{equ:RG2eq}),
	    the simulation results (dots) follow $\langle R_G^2\rangle_\tau^{(eq)} \simeq \alpha N^2/12$~ (dashed line).}
  \label{fig:RG2ofN}
\end{figure}

It is defined as
\begin{align}
  R_G^2 &= \sum_{kl} d(k,l)^2 p_k p_l,\label{equ:RG}
\end{align}
with $d(k,l)$ denoting the distance between sites $k$ and $l$.
Thus, the gyration radius gives the mean squared distance between the two diffusion particles.
In a linear lattice with closed boundary conditions $d(k,l) = |k-l|$.
In the case of a ring of size $N$ the distance is given by
\begin{align}
  d(k,l) &= \begin{cases}
	      |k-l|, &\text{if~} |k-l| \leq N/2\\
	      N-|k-l|, &\text{else.}
            \end{cases}
\end{align}

We say a state is localized if the radius of gyration is small.
With the definition of the matrix $R \in \mathbb{R}^{N \times N}$ with the components $R_{kl} = d(k,l)^2$
we find:
\begin{align}
   R_G^2 &= \vec{p}^T R \vec{p}.
\end{align}
\begin{figure}
  \includegraphics[width=\linewidth]{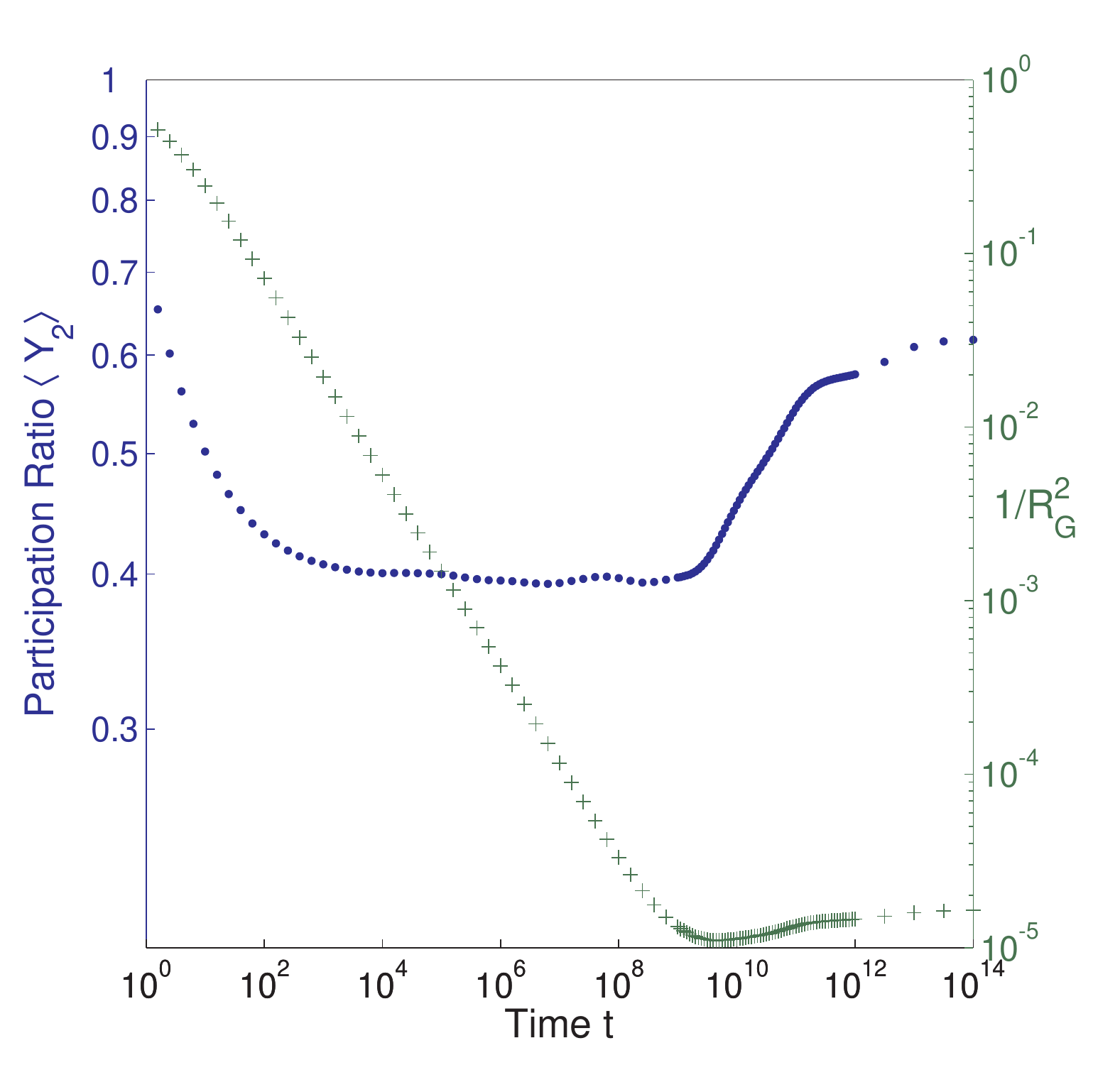}
  \caption{Temporal evolution of $\langle Y_2\rangle_{p_0, \tau}$ (\textbullet) in comparison to $1/\langle R_G^2\rangle_{p_0, \tau}$ ({\bf +}),
  calculated as average over 100 landscape realizations of size $N = 11^3$ and $\alpha = 0.37$.
	}
  \label{fig:RG2temporal}
\end{figure}
This has the same structure as $Y_2 = \vec{p}^T \mathbb{I} ~\vec{p}$ with $\mathbb{I}$ being the identity matrix. But, whereas
the bilinear form $\mathbb{I}$ is a diagonal matrix, $R$ has vanishing diagonal entries and is dominated by non-diagonal ones. In analogy to
the calculations that led to Eq. (\ref{equ:pTpAveraged}), we find:
\begin{align}
  \langle R_G^2 \rangle_{p_0} &= \frac{1}{N} \sum_{ij} \left( \vec{Z}_i^T \vec{Z}_j \right) \left( \vec{X}_i^T R \vec{X}_j\right) e^{-(\lambda_i + \lambda_j)t}.
  \label{equ:RG2Averaged}
\end{align}
The equilibrium value in disorder average can be computed easily:
\begin{align}
  \langle R_G^2 \rangle_\tau^{(eq)} &= \sum_{kl} d(k,l)^2 \langle p_k p_l \rangle_\tau^{(eq)}\\
    &= \langle p_k p_l \rangle_{\tau,k \neq l}^{(eq)} \sum_{kl} d(k,l)^2
\end{align}
The two factors can now be evaluated independently:
\begin{align}
  &\left\langle p_k p_l \right\rangle_{\tau, k\neq l}^{(eq)} = \frac{1}{2} \left\langle \left( p_k + p_l\right)^2 - p_k^2 - p_l^2 \right\rangle_{\tau, k\neq l}^{(eq)}\\
      &= \frac{1}{2}\left\langle \left( p_k + p_l\right)^2\right\rangle_{\tau, k\neq l}^{(eq)} - \left\langle p^2 \right\rangle_{\tau}^{(eq)}
      ~=~ \frac{\alpha}{N^2},
\end{align}
where we have used the results of \cite{Sokolov2010}.
The evaluation of $\sum_{k,l=1}^N d(k,l)^2$ yields
\begin{align}
  \sum_{k,l=1}^N d(k,l)^2 \simeq \frac{N^4}{12},
\end{align}
so that
\begin{align}
  \langle R_G^2 \rangle_\tau^{(eq)} &\simeq \frac{\alpha N^2}{12}
  \label{equ:RG2eq}
\end{align}
Fig. \ref{fig:RG2ofN} shows the results of simulations of Eq. (\ref{equ:RG2Averaged}) in the limit $t \rightarrow \infty$ for different $\alpha < 1$ and
system sizes $N$ and the comparison to Eq. (\ref{equ:RG2eq}).
\begin{figure}[t!]
    \centering{\includegraphics[width=\linewidth]{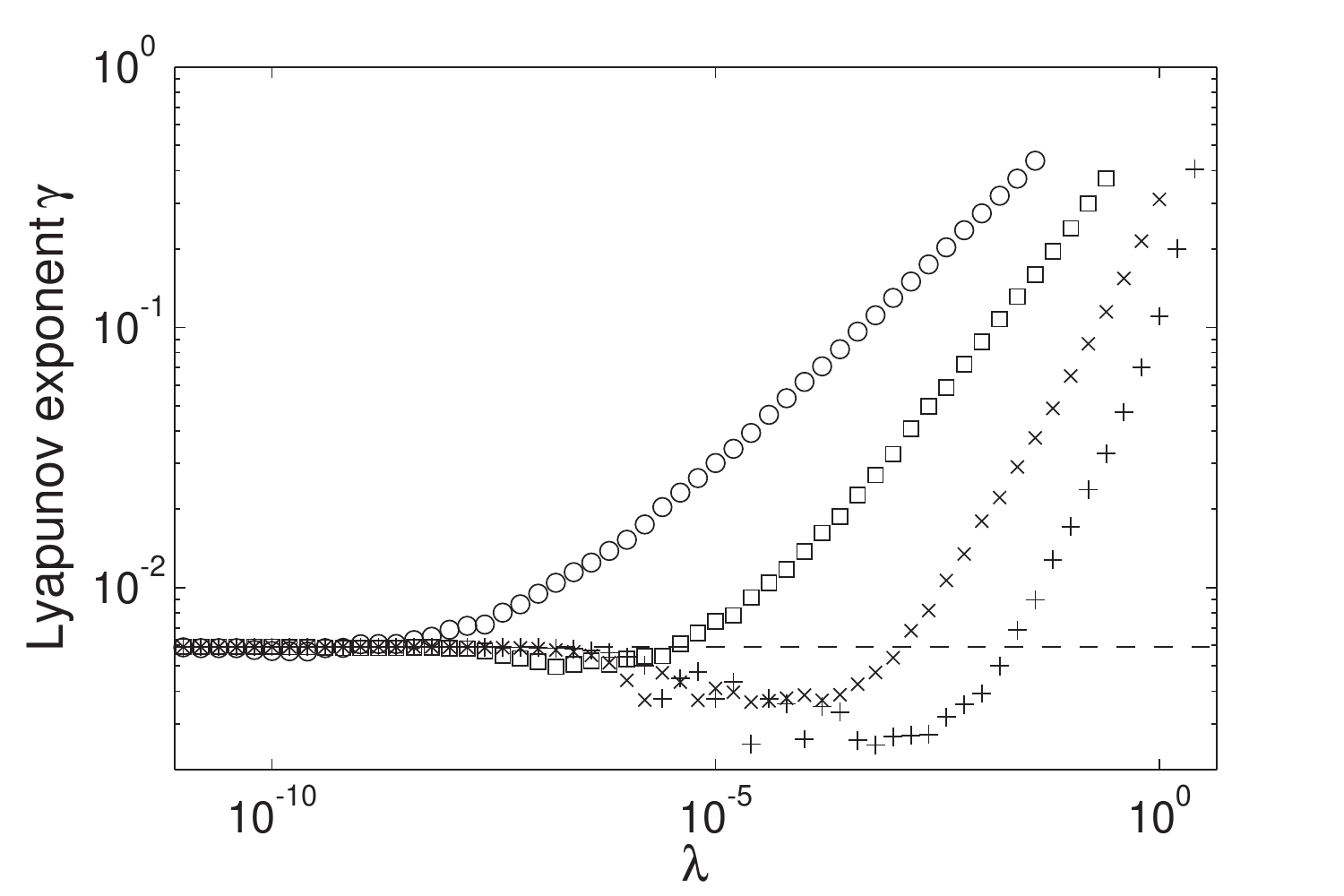}}
    \caption{Lyapunov exponent $\gamma$ for $\alpha = 0.5$ ($\circ$),  $\alpha = 0.75$ ($\square$), $\alpha = 1.25$ ($\times$), and $\alpha = 2$ ({\bf +}) in comparison
    to $\lambda$ for $N = 1331$, averaged over 100 landscape realizations.}
    \label{fig:gammaOfECloseUp}
\end{figure}

The important information is that the equilibrium value of the radius of gyration grows with the system size whereas the equilibrium value of $\langle Y_2 \rangle_\tau$
becomes independent of $N$ (for $N$ large enough), i.e. the states localized according to the $Y_2$ criterion are delocalized with respect to their gyration radius. 

The reason is quite clear. If a site attracts a considerable amount of probability in the equilibrium state $\vec{p}^{(eq)} = (\tau_1, \tau_2, \ldots , \tau_N)^T/\sum_k\tau_k$,
let us call it a ``deep trap''. If there is only one deep trap in the landscape, $Y_2$ is large, and $R_G$ is small. Many landscape realizations do contain \emph{more than one} deep trap,
in which case $Y_2$ is still large, but $R_G$ might be large as well, being of the size of the distance between the deep traps. 
The expectation of the distance between the deep traps grows with the system size and thus the radius of gyration does also grow, whereas
the participation ratio is independent of the deep traps' positions.

Apart from the equilibrium value, we can also compare the temporal evolution of $1/\langle R_G^2\rangle_{p_0, \tau}$ to the one of $\langle Y_2\rangle_{p_0, \tau}$.
Fig. \ref{fig:RG2temporal} depicts such a comparison with numerical averages
over 100 landscape realizations.

Altogether, the phenomenon of dynamical localization is pertinent to the behavior of one, very specific localization measure. In the range of times when the 
dynamical localization is observed, the probability that the two independent random walkers starting at the same site appear to be at the same site at a later 
instant of time $t$ (as represented by $Y_2$) stagnates, but the mean squared distance between them (given by $R_G$) grows continuously. $Y_2$ is thus a measure of the
concentration of the state on some small subset of sites; $R_G$, on the other hand, quantifies the size of the spatial domain in which the particles are confined. 
Both the dynamically localized state and the equilibrium state of the system are essentially delocalized with respect to this measure.

\subsection{Localization of the eigenstates}
\begin{figure}
  \includegraphics[width=\linewidth]{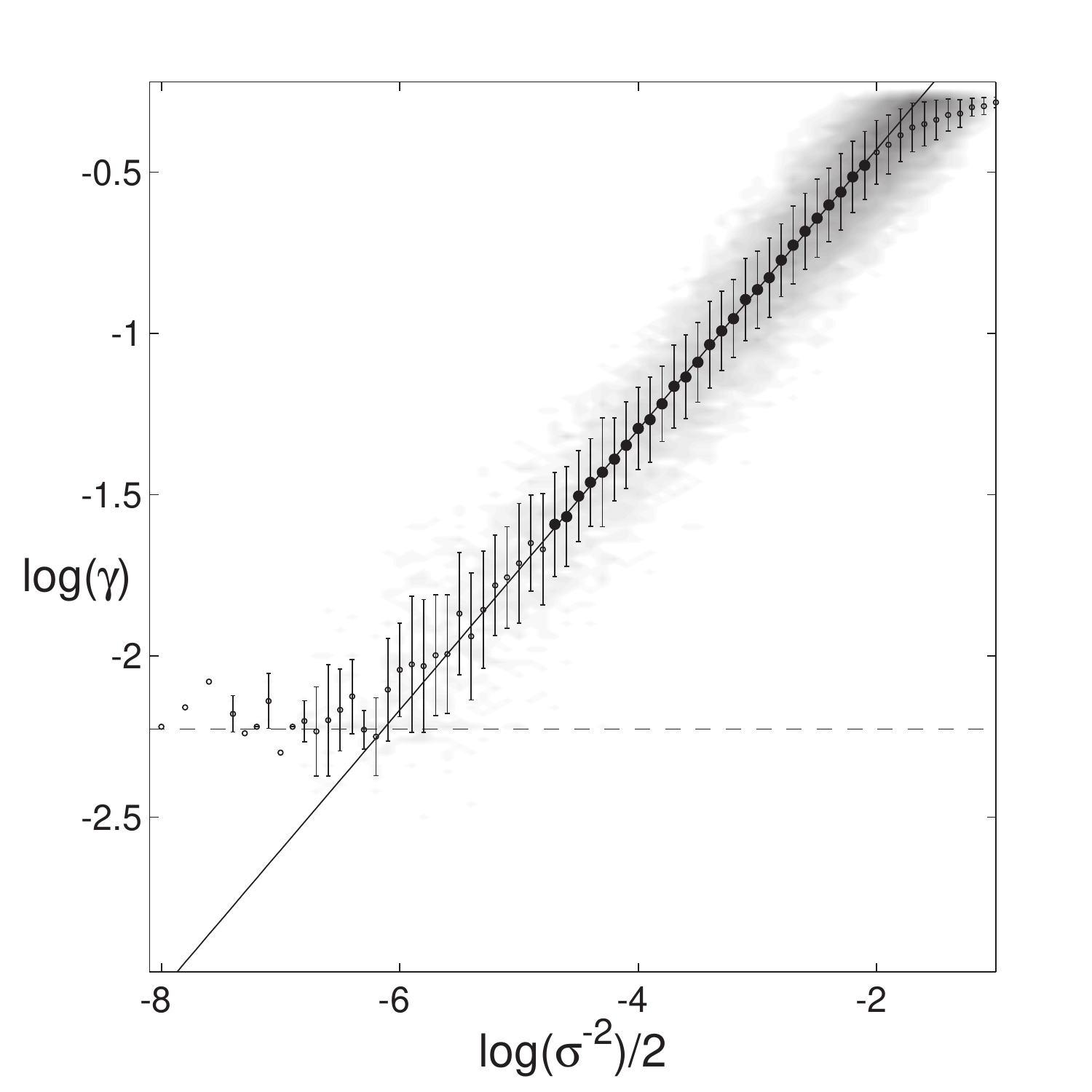}
  \caption{The Lyapunov exponent $\gamma$ vs. inverse radius of gyration $\sqrt{\sigma^{-2}}$ (see Eq. (\ref{equ:sigma-2})) on double logarithmic scales.
      The shaded area gives the density of single data points, whereas the dots represent averaged values. The larger circles define the
      ensemble for a linear fit (solid line), yielding $\log \gamma = 0.440 + 0.434 \cdot \log \sqrt{\sigma^{-2}}$.
      The dashed line represents the numerical zero $\log(2N)/N$.}
  \label{fig:gammaOfsigma}
\end{figure}

The equilibrium state is essentially an eigenvector of $L$ to its zero eigenvalue; therefore it is interesting to look at the eigenvalue localization
properties of $L$ also for other eigenvalues $\lambda$. The corresponding measure of localization is given by 
the Lyapunov exponent $\gamma$ which is defined as follows:
Let $\pi_k$ be the components of an eigenvector of $L$ to eigenvalue $\lambda$. Then the following relation holds:
\begin{align}
    \begin{pmatrix}
       \frac{\pi_k}{\tau_k}\pb{2em}\\ \frac{\pi_{k+1}}{\tau_{k+1}}\pb{2em}
    \end{pmatrix}
    &= \underbrace{\begin{pmatrix}
	  0 & 1\\ -1 & 2 - \lambda \tau_k
       \end{pmatrix}}_{=~T_k}
    \begin{pmatrix}
	\frac{\pi_{k-1}}{\tau_{k-1}}\pb{2em}\\ \frac{\pi_k}{\tau_k}\pb{2em}
    \end{pmatrix}.
    \label{equ:Tk2}
\end{align}
The matrix $T_k$ which transfers between the vector on the r.h.s. and the vector on the l.h.s. is called the $k$th transfer matrix.
The Lyapunov exponent is now defined as
\begin{align}
     \gamma = \lim_{N \rightarrow \infty} \frac{1}{N} \left\langle \log \| T_N \ldots T_2 T_1 \|_1\right\rangle_\tau,
     \label{equ:Defgamma}
\end{align}
with $\| M \|_1 = \max_{j = 1,2} \left( |m_{1j}| + |m_{2j}| \right)$ denoting the maximum absolute column sum of the matrix $M$.
Eq. (\ref{equ:Defgamma}) together with Eq. (\ref{equ:Tk2}) defines $\gamma(\lambda)$ regardless of the fact whether $\lambda$ is an eigenvalue of $L$ or not.
(Note that the vectors in Eq. (\ref{equ:Tk2}) are actually built from the components of the eigenvectors of $L^T$.
In the limit $N \rightarrow \infty$, however, it can be shown that the Lyapunov exponent $\gamma(\lambda)$
of both $L$ and $L^T$ is the same.)

As the Lyapunov exponent gives the rate of exponential growth in an (eigen-)vector, it is a standard measure for localization: $\gamma = 0$ means that
the state is delocalized whereas a high Lyapunov exponent means strong localization. Considering this, it seems peculiar to us that the equilibrium state
of the trap model should have a high value of participation ratio but a vanishing Lyapunov exponent $\gamma$:
\begin{figure}
  \includegraphics[width=\linewidth]{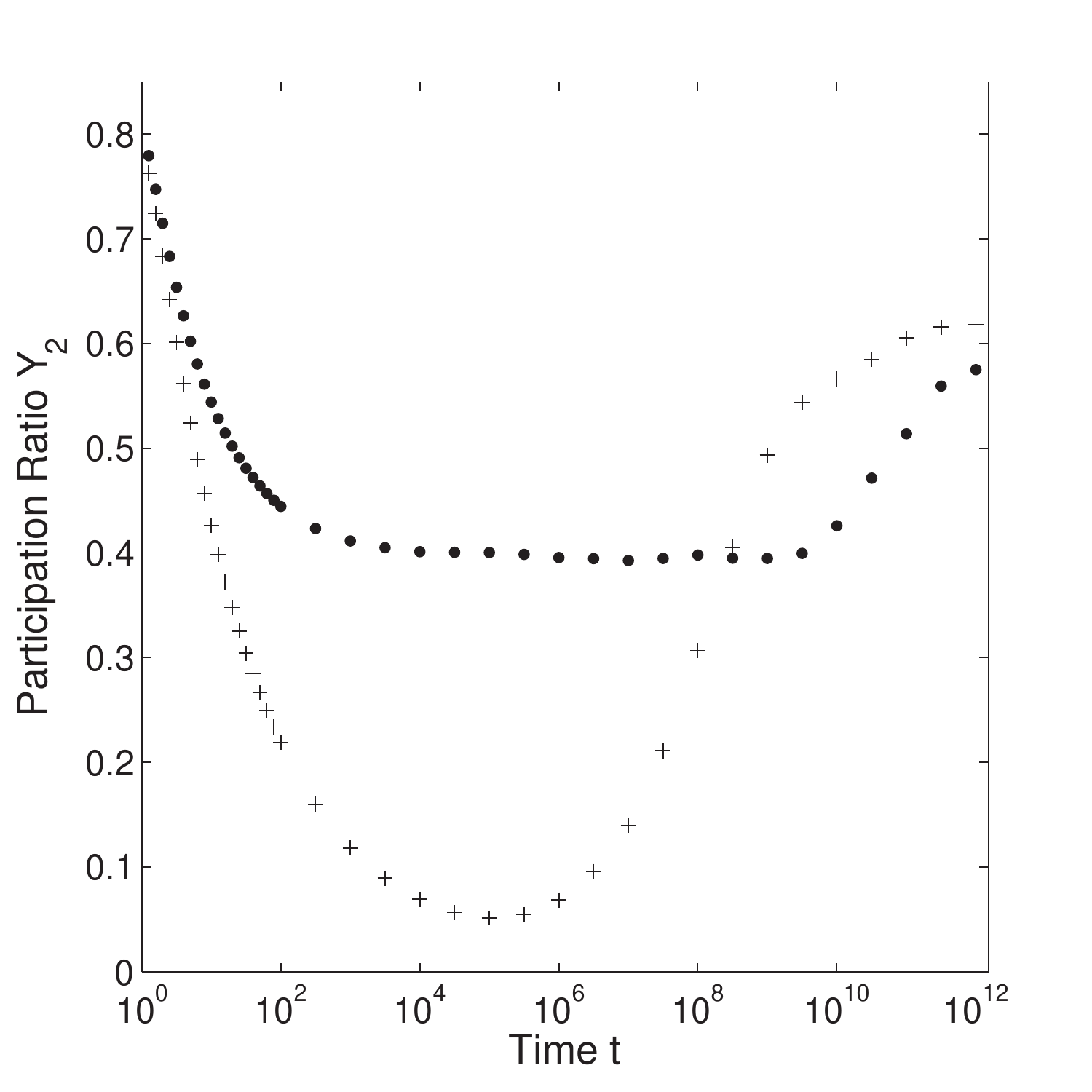}
  \caption{Participation ratio as a function of time for 1D and 3D systems. The parameters are: $N = 11^3$, $\alpha = 0.37$, the
  number of realizations is 100.}
  \label{fig:Y21D3D}
\end{figure}
\begin{align}
  \gamma(0) &= \lim_{N \rightarrow \infty} \frac{1}{N} \left\langle \log \left\| \begin{pmatrix}
	  0 & 1\\ -1 & 2
       \end{pmatrix}^N \right\|_1\right\rangle_\tau\\
       &= \lim_{N \rightarrow \infty} \frac{1}{N} \log \left\| \begin{pmatrix}
	  1-N & N\\ -N & N+1
       \end{pmatrix}\right\|_1\\
       &= 0.
\end{align}
Thus, the equilibrium state is essentially delocalized with respect to the its Lyapunov exponent.
Since in all our simulations we consider finite systems, the measured Lyapunov exponent at $\lambda = 0$ is of order $\log(2N)/N$. 

A double-logarithmic plot of the Lyapunov exponent for $N = 1331$ and $\alpha \in \{ 0.5, 0.75, 1.25, 2 \}$
is shown in Fig. \ref{fig:gammaOfECloseUp}. For all four values of $\alpha$ the Lyapunov exponent $\gamma(0)$ is of order $\log(2N)/N$
(dashed line) which corresponds to zero in the infinite system size limit.
\begin{figure*}
  \includegraphics[width=.48\linewidth]{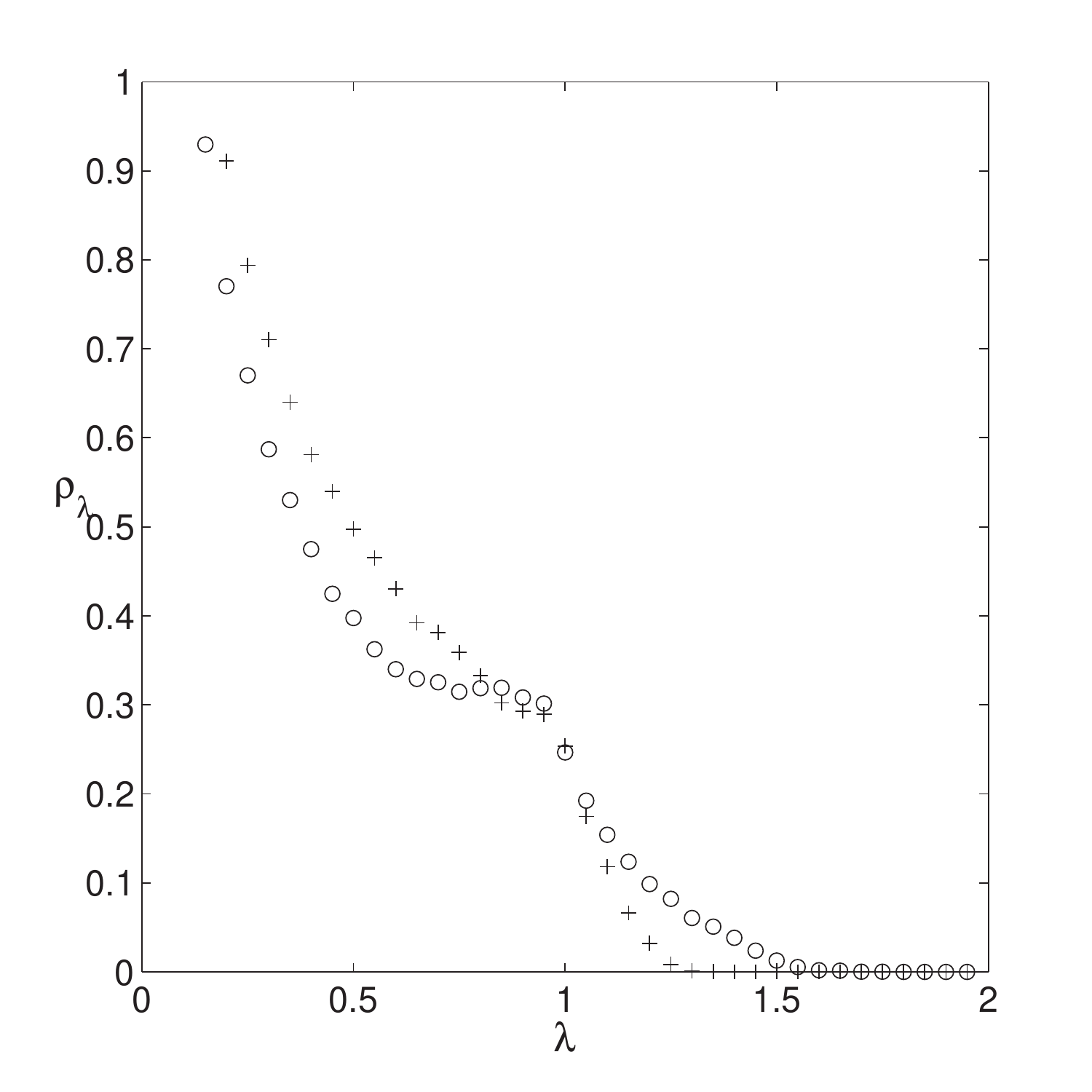}
  \includegraphics[width=.48\linewidth]{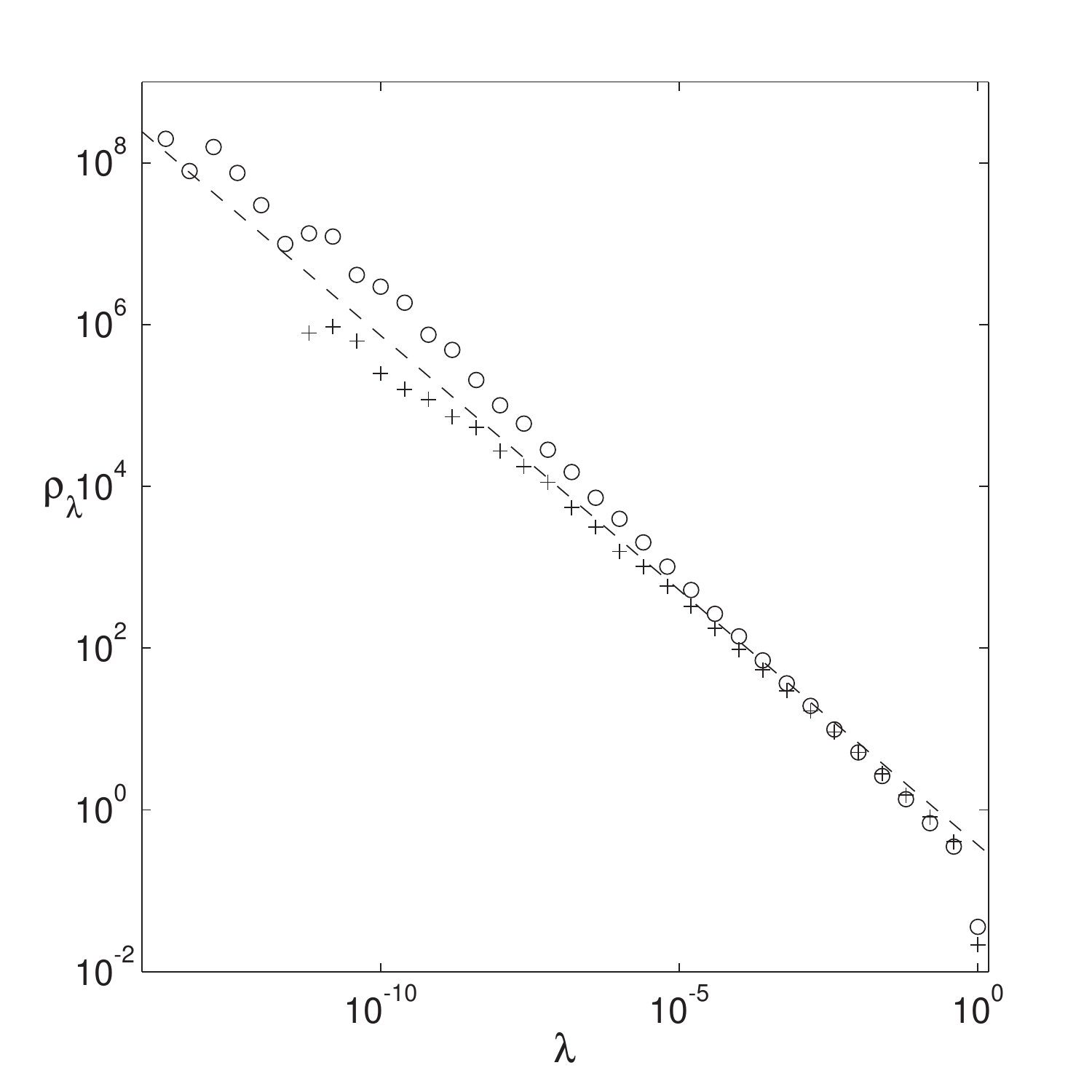}
  \caption{Spectrum of $L$ ($\circ$) and $L^{(3D)}$ ({\bf +}) in both linear and double-logarithmic scale with system size $N = 1331$ and Pareto-exponent $\alpha = 0.37$. The
    spectrum has been averaged over 100 landscape realizations.
    The dashed line in the double-logarithmic plot is $\rho_\lambda = \alpha\lambda^{\alpha-1}$ which is a theoretical result
    for the infinite-dimensional REM-like case \cite{Bovier2005} (not a fit)
    practically coinciding with the 3D data.}
  \label{fig:spectrum}
\end{figure*}

The Lyapunov criterion of localization gives information which is strongly correlated with the one delivered by the gyration radius of single eigenstates. 
The inverse squared gyration radius for an eigenstate can be defined via
\begin{align}
  \sigma_k^{-2} = \left( \frac{\sum_{i,j=1}^N d(i,j)^2 X_{ki}^2 X_{kj}^2}{\sqrt{\sum_{j=1}^N X_{kj}^4}}\right)^{-1}
  \label{equ:sigma-2},
\end{align}
and $X_{kj}$ being the $j$th entry of the $k$th eigenvector. In Fig. \ref{fig:gammaOfsigma} we plot $\gamma_k$ vs. $\sqrt{\sigma_k^{-2}}$ on double
logarithmic scales. Strong correlation (linear on double logarithmic scales, i.e. a power-law dependence) between both is seen. 
A linear fit shows that $\log \gamma = 0.440 + 0.434 \cdot \log \sqrt{\sigma^{-2}}$. This means that the Lyapunov exponent and the gyration radius provide related information on
the eigenstate localization: the eigenstates confined to a smaller region typically decay faster outside of it, and vice versa.

\subsection{Spectral properties and dynamical localization}
\label{subsec:spectrum}

Now we turn to another important question, namely where exactly the information on the dynamical localization is hidden: Is it determined by the spectral
properties of the Laplacian matrix, or is it mostly coded in the entries of the matrix $G$ composed from the eigenvectors of $L$?
To understand this we compare the behavior of a one-dimensional system of $N$ sites with the one of its three-dimensional analogue, where dynamical localization is absent
\cite{Fontes1999}.
The total number of sites shall also be $N$, i.e. the side length of the cube is $N^{1/3}$. The rates for the 3D case are
as follows:
\begin{align}
    w_{k\rightarrow l}^{(3)} &= \begin{cases}
                          -1/\tau_k, &\text{if~} k = l,\\
                          1/(6\tau_k), &\text{if~} k,l \text{~neighbors},\\
                          0, &\text{else}.
                        \end{cases}
\end{align}
A comparison between the disorder averaged evolution of $Y_2$ on the one- and three-dimensional case is shown in Fig. \ref{fig:Y21D3D}.
As Fontes \emph{et al.} \cite{Fontes1999} proved, there is no dynamical localization in the 3D case, i.e. $\lim_{t\rightarrow \infty} \lim_{N \rightarrow \infty} \langle Y_2\rangle_\tau = 0$,
which, for finite lattices
with site number $N$, can be understood as $\lim_{N \rightarrow \infty} \min_t \langle Y_2\rangle_\tau = 0$.
In Fig. \ref{fig:Y21D3D} this can be seen by the fact that $\min_t\langle Y_2^{3D}\rangle_\tau$
is of the order of $N^{-1}$ and $\min_t\langle Y_2^{1D}\rangle_\tau$ is considerably larger although in both cases the number of sites is $N = 11^3$.

In Fig. \ref{fig:spectrum} we plotted the spectrum $\rho_\lambda$ for both Laplacians $L$ and $L^{(3D)}$ on linear and on double-logarithmic scales. Note that the cusp at $\lambda = 1$ is
due to the fact that we use pure Pareto-distributed waiting times and it does not show up for analytic distributions as for
example a one-sided L\'evy-distribution with corresponding asymptotics. In any case, the behavior of $\rho_\lambda$ for large values of $\lambda$ does not seem interesting to us because
we are looking for hints for the long-time behavior of the system which means we are concerned about small $\lambda$. The double-logarithmic plot of Fig. \ref{fig:spectrum} reveals that for
small values of $\lambda$, the spectrum $\rho_\lambda$ exhibits a power-law behavior. 
The exponent in the 3D case is in fact equal to $\alpha - 1$, as the dashed line in the double-logarithmic plot indicates. This is the exponent Bovier and Faggionato \cite{Bovier2005}
determined for infinite-dimensional models. Note, that the dashed line in Fig. \ref{fig:spectrum} is not a fit. The absolute value of the exponent for the 1D case is slightly larger than in 3D.
\begin{figure}
  \includegraphics[width=\linewidth]{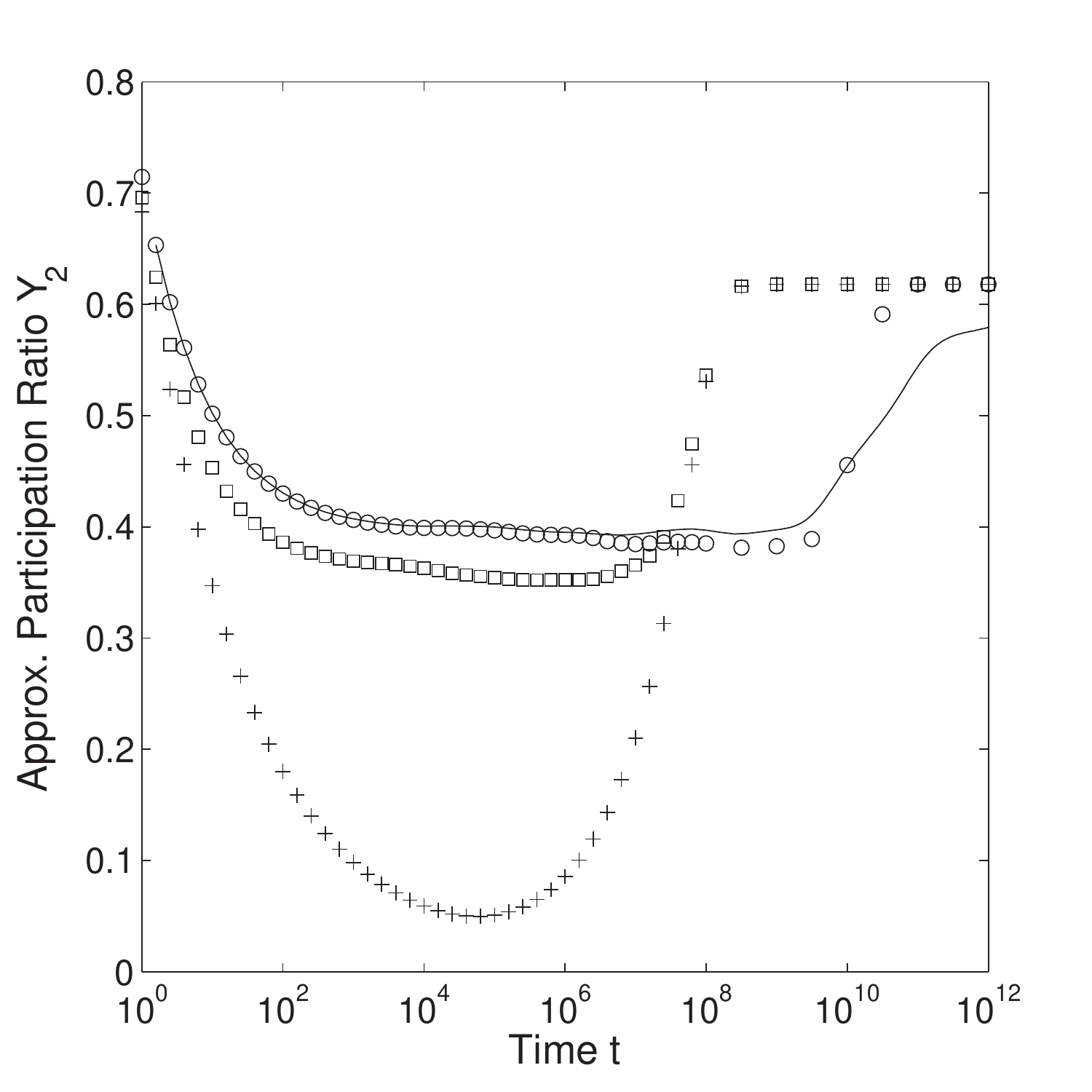}
  \caption{Approximation of Eqs. (\ref{equ:Factorized}) and (\ref{equ:1Dwith3Deigenvalues}) for $N = 11^3$, $\alpha = 0.37$ and 100 landscape realizations.
	    The exact result for the 1D case is given by the solid line and the approximation of Eq. (\ref{equ:Factorized}) is depicted by circles, the 3D case by plus signs. The approximation of Eq. (\ref{equ:1Dwith3Deigenvalues})
	    is represented by squares.}
  \label{fig:Factorized}
\end{figure}
Nevertheless, the qualitative behavior of both the spectrum of $L$ and $L^{(3D)}$ does not differ strongly.

The question is, whether the different exponents of $\rho_\lambda$
do contribute to the effect of dynamical localization, or whether they are only an expression of the different time scales of the corresponding random walks.
In order to investigate this, as well as the interplay between the eigenvectors and eigenvalues, we test the following decoupling approximation of Eq. (\ref{equ:pTpAveraged}):
\begin{align}
   \langle Y_2\rangle_{p_0, \tau} &= \sum_{ij} \left\langle G_{ij} e^{-(\lambda_i + \lambda_j)t} \right\rangle_\tau\nonumber\\
   &\simeq \sum_{ij} \left\langle G_{ij}\right\rangle_\tau \left(e^{-\tilde \lambda_i}\right)^t \left(e^{-\tilde \lambda_j}\right)^t = \langle \tilde{Y}_2\rangle_{p_0, \tau}.
   \label{equ:Factorized}
\end{align}
with $e^{-\tilde \lambda_i} = \langle e^{-\lambda_i} \rangle_\tau$.
The result of this approximation is depicted in Fig. \ref{fig:Factorized}.
The figure shows that the decoupling approximation adequately reproduces the results of direct simulations
and that the correlations between eigenvalues and the entries of $G$ do not actually play a role.
\begin{figure}
  \begin{center}
  \includegraphics[width=.9\linewidth]{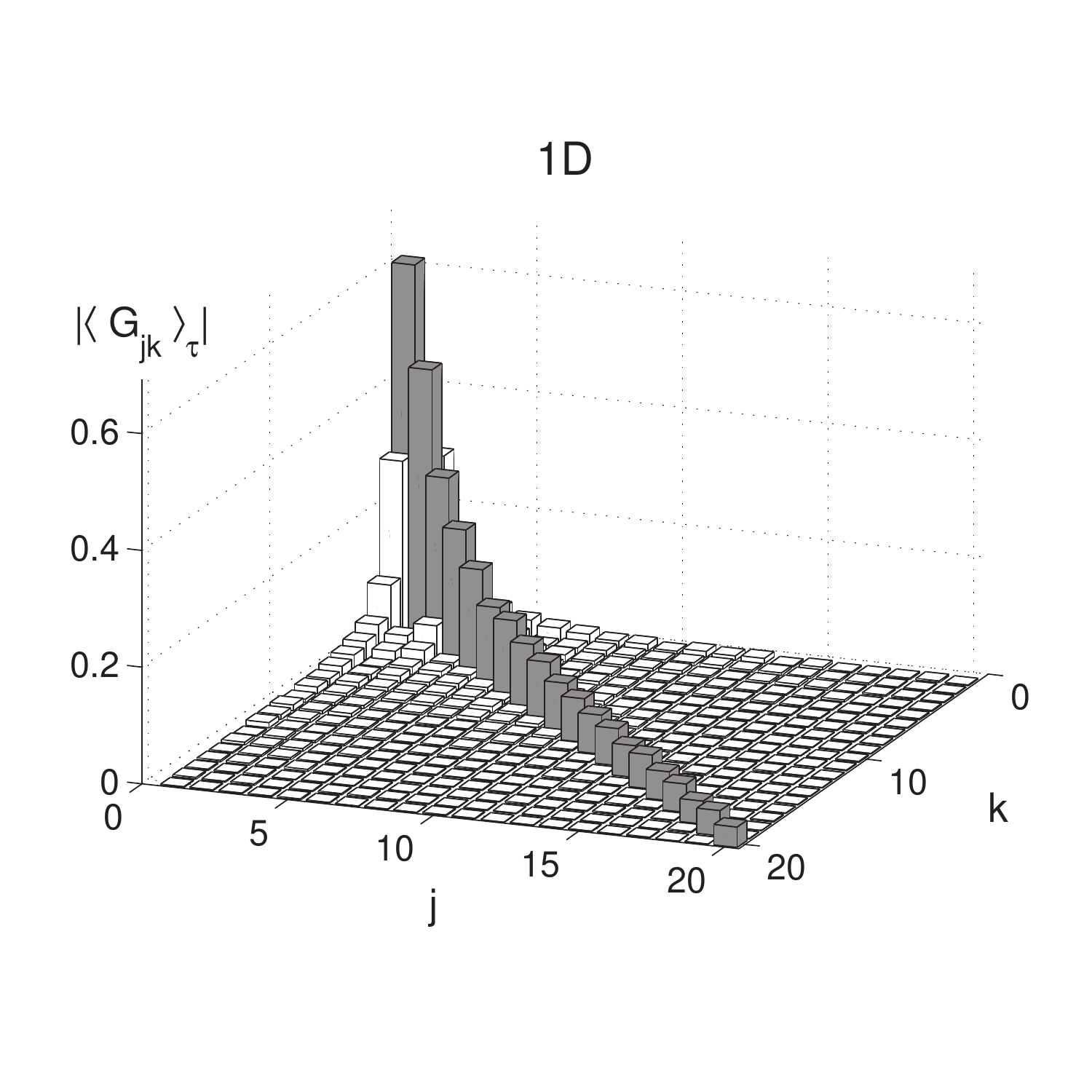}
  \includegraphics[width=.9\linewidth]{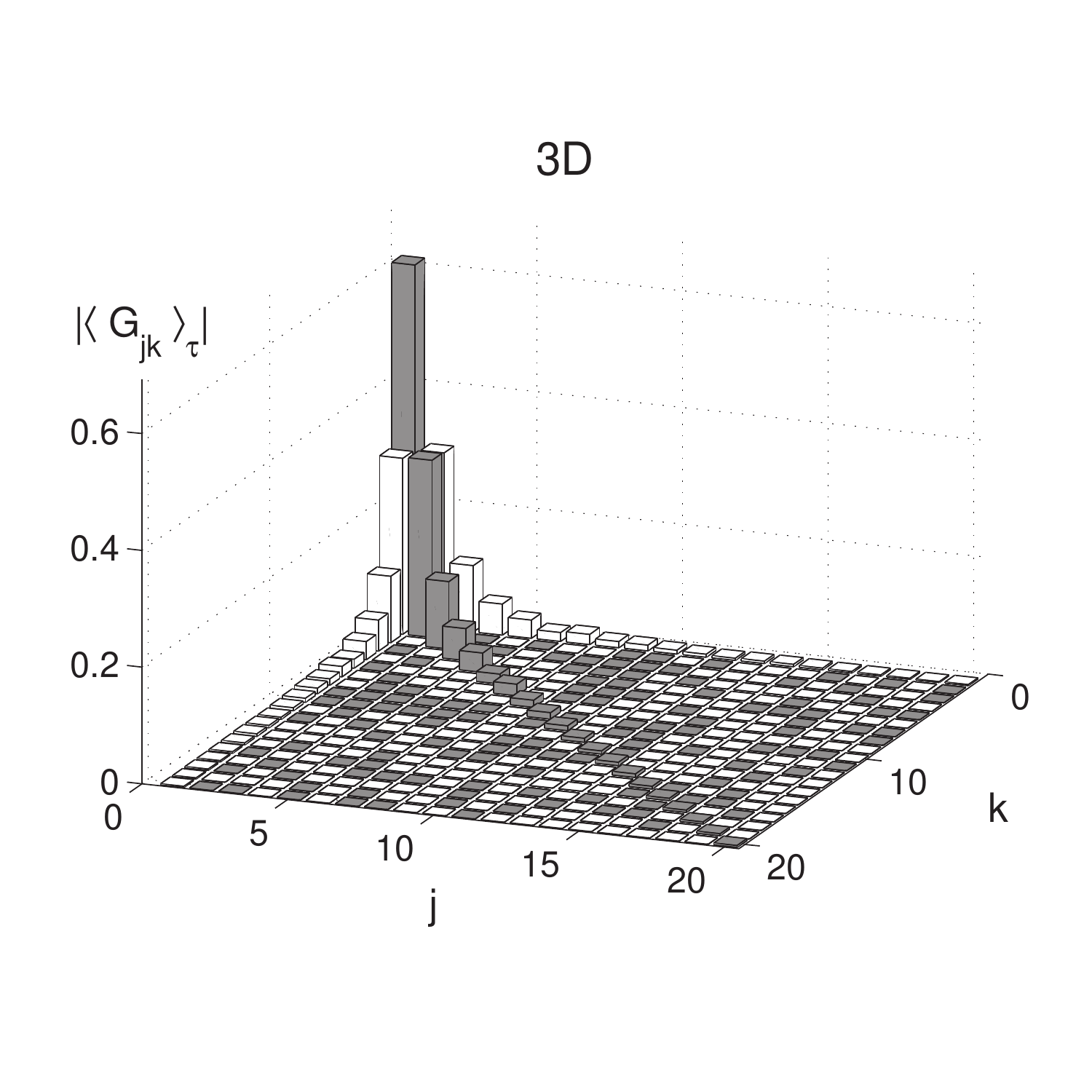}
  \end{center}
  \caption{First $20 \times 20$ entries of the matrix $G$ defined by Eq. (\ref{equ:DefG}). $N = 1331$, $\alpha = 0.37$, 100 realizations. Gray bars represent positive
	    entries and white bars negative ones.}
  \label{fig:GAveraged}
\end{figure}

To elucidate the the influence of the different spectra in the 1D and 3D case, we simply substitute
the eigenvalues of the 1D case by the ones of the 3D case (i.e. use 1D $G$-matrix together with the corresponding 3D eigenvalues):
\begin{align}
   \langle \tilde{Y}_2^{(1D)}\rangle_{p_0, \tau} &\simeq \sum_{ij} \left\langle G^{(1)}_{ij}\right\rangle_\tau \left( e^{-\tilde{\lambda}^{(3)}_i}\right)^t \left( e^{-\tilde{\lambda}^{(3)}_j}\right)^t,
   \label{equ:1Dwith3Deigenvalues}
   \intertext{but still take}
   \langle \tilde{Y}_2^{(3D)}\rangle_{p_0, \tau} &\simeq \sum_{ij} \left\langle G^{(3)}_{ij}\right\rangle_\tau \left( e^{-\tilde{\lambda}^{(3)}_i}\right)^t \left( e^{-\tilde{\lambda}^{(3)}_j}\right)^t.\nonumber
\end{align}
Thus, in this approximation, the only difference between the 1D and the 3D case is the shape of the matrix $G$ defined by Eq. (\ref{equ:DefG}). The result of the approximation Eq. (\ref{equ:1Dwith3Deigenvalues})
is also presented in Fig. \ref{fig:Factorized}. Although the plateau of the case Eq. (\ref{equ:1Dwith3Deigenvalues}) lies lower than the one of Eq. (\ref{equ:Factorized}),
it is still there and pronounced. Thus, the exact form of the spectrum is only responsible for the height and the duration of the plateau, not for the fact of its presence or absence. 
We thus conclude that the difference in the spectrum $\rho_\lambda$ is primarily responsible for the different time
scales but \emph{not} for the fact of dynamical localization.

Since we have shown that it is the matrix $G$ built from the eigenvectors of $L$ which codes for the presence or absence of dynamical localization, the question arises,
what properties of its elements are mainly responsible for it. The elements of $G$ differ strongly in their magnitude. 
Is the plateau visible in Fig. \ref{fig:Factorized} the result of a complex interplay between positive and negative small components of $G$ or
is its presence dominated by the behavior of large components of $G$ other than $G_{11}$ (coding for the equilibrium)?

To attack this question, the first $20 \times 20$ entries of $\langle G\rangle_\tau$ in the 1D and 3D case are depicted in Fig. \ref{fig:GAveraged}.
In order to make the important information
visible, the $z$-axis represents the absolute values of the elements, $|\left\langle G_{jk}\right\rangle_\tau|$, whereas the sign of the entries is indicated by the color of the bars: Gray bars represent positive
entries and white bars negative ones. Since the equilibrium state is the same in 1D and 3D, $\langle G_{11} \rangle = 1-\alpha$ in both cases. In the 3D case however,
the values of the other diagonal entries decay more rapidly than in the 1D case. Moreover, while in the 3D case, for $k \neq j$ and $j,k > 1$,
the entries $\left\langle G_{jk}\right\rangle_\tau$ are close to zero and of fluctuating sign, in 1D case all shown off-diagonal entries are negative (although for
larger indices $j,k$ they also fluctuate).

\begin{figure}
  \includegraphics[width=1.1\linewidth]{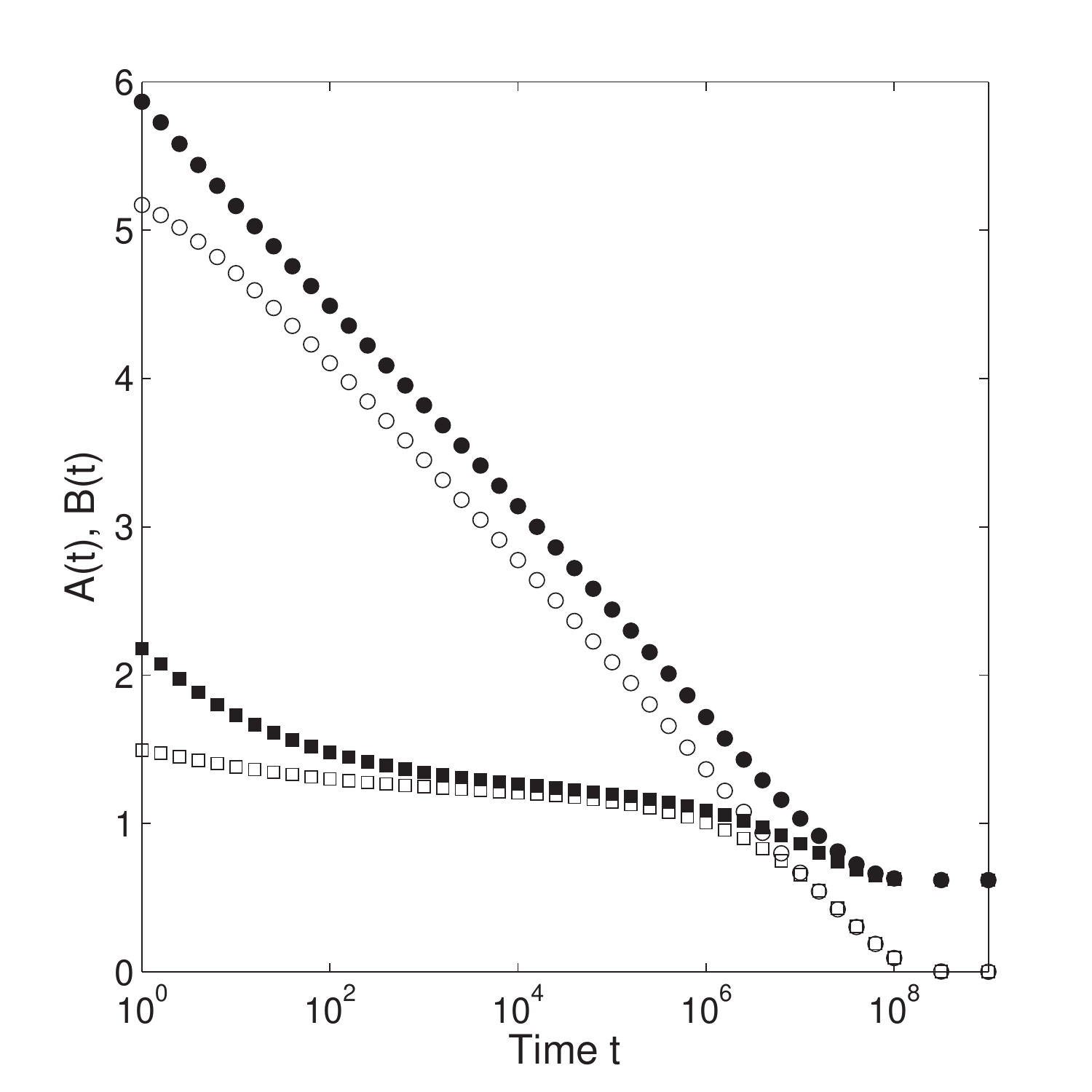}
  \caption{Temporal evolution of the summands $A$ (filled symbols) and $B$ (empty symbols) of Eq. (\ref{equ:SplittedSum}) for the 1D (circles) and 3D (squares) case.
   Simulations with system size $N = 1331$, $\alpha = 0.37$, and 100 landscape
    realizations.}
  \label{fig:Factorization}
\end{figure}

Let us split the diagonal and the non-diagonal contributions to the overall sum over eigenstates: 
\begin{align}
  \langle Y_2 \rangle_{p_0, \tau} &= \underbrace{\sum_{j} \left\langle G_{jj}\right\rangle_\tau \left\langle e^{-\lambda_j}\right\rangle_\tau^{2t}}_{= A}\nonumber\\
	&\quad + \underbrace{\sum_{j\neq k} \left\langle G_{jk}\right\rangle_\tau \left\langle e^{-\lambda_j}\right\rangle_\tau^t \left\langle e^{-\lambda_k}\right\rangle_\tau^t}_{= -B}
	\label{equ:SplittedSum}
\end{align}
The diagonal elements of $G$ are always positive, and so is $A$. According to the simulations, the term $-B$ is negative such that $B$ is positive. 
Fig. \ref{fig:Factorization} shows the temporal evolution of $A$ and $B$ in both 1D and 3D case, where in both cases we have used
again the spectrum of the 3D system. It seems that the plateau in Fig. \ref{fig:Factorized} is indeed the result of an interplay of the positive diagonal entries in $A$ and
the mostly negative ones in $-B$. In 1D both terms decay logarithmically but the sum $A-B$ stays constant over a long time. In three dimensions the terms behave completely different.
This shows that the presence or absence of dynamical localization is fully coded in the matrix $G$, i.e. depends on the properties of eigenvectors of $L$, and not so much on its spectrum.
It is strongly dominated by the scalar products of eigenvectors to different eigenvalues (which are not orthogonal since $L$ is not symmetric) entering the non-diagonal terms,
and in this sense indeed has to do with eigenvector localization, although the direct relation is not clear yet. 

\section{Conclusions}

In the present work we applied an algebraic approach to investigate the phenomenon of dynamical localization in the random trap model with
power-law distributed mean waiting times.
Apart from its formal elegance, the approach works extremely well as a computational scheme,
since it allows for obtaining numerically exact results for the participation ratio $Y_2$ averaged over initial conditions and thermal histories
in a given system's realization. Only a single averaging procedure over the samples is necessary.
Because this approach enables us to compute the relevant properties at any given time, it allows for considering longer observation times than in the 
direct Monte-Carlo simulations.

The phenomenon of dynamical localization, as observed in the participation ratio, is a very peculiar property of one-dimensional trap models.
Its physical interpretation is that the probability, that two particles with the same starting position are found together, is surprisingly high for intermediate time scales.
In this context, \emph{intermediate time scales} means that full equilibration is not yet established.
The fact, that the particles often meet at the same site, does not mean that they stay close to each other: the typical
distance between the particles continuously grows with time until the terminal equilibration is reached. Moreover, the equilibrium state, which is the strongest localized one
with respect to $Y_2$, is essentially delocalized with respect to its radius of gyration and its Lyapunov exponent.
Numerical simulations show that the later two
are well-correlating localization measures for the eigenstates of the system. 

We have moreover shown, that the phenomenon of dynamical localization is only marginally connected with the spectral properties of the Laplacian operator governing
the system's dynamics, and is dominated by the properties of the eigenfunctions of $L$, which outlines the direction of further investigations.  

\section{Acknowledgements}
The financial support of DFG within the IRTG 1740 ``Dynamical Phenomena in Complex Networks: Fundamentals and Applications'' is gratefully acknowledged.

Furthermore, the authors thank Prof. D. H. U. Marchetti for useful discussions.

\end{document}